\documentclass[12pt]{article}
\usepackage{jheppub}

\DeclareUnicodeCharacter{2212}{-} \DeclareUnicodeCharacter{FB01}{-} \DeclareUnicodeCharacter{FB02}{-}

\usepackage{amsmath,bbm,array,cleveref,amsfonts,xcolor,subcaption,graphicx,wrapfig,lscape,youngtab,bbold,float,multirow,longtable,rotating,epstopdf,breqn}
\usepackage[export]{adjustbox}
\newlength{\overwritelength}
\newlength{\minimumoverwritelength}
\newcommand{\overwrite}[3][red]{%
  \settowidth{\overwritelength}{$#2$}%
  \ifdim\overwritelength<\minimumoverwritelength%
    \setlength{\overwritelength}{\minimumoverwritelength}\fi%
  \stackrel
    {%
      \begin{minipage}{\overwritelength}%
        \color{#1}\centering\small #3\\%
        \rule{1pt}{9pt}%
      \end{minipage}}
    {\colorbox{#1!50}{\color{black}$\displaystyle#2$}}}

\newcommand{\be}{\begin{equation}}
\newcommand{\ee}{\end{equation}}
\newcommand{\beq}{\begin{equation}}
\newcommand{\beql}[1]{\begin{equation}\label{#1}}
\newcommand{\eeq}{\end{equation}}
\newcommand{\ba}{\begin{array}}
\newcommand{\ea}{\end{array}}
\newcommand{\bea}{\begin{eqnarray}}
\newcommand{\beal}[1]{\begin{eqnarray}\label{#1}}
\newcommand{\eea}{\end{eqnarray}}
\newcommand{\ben}{\begin{enumerate}}
\newcommand{\een}{\end{enumerate}}
\newcommand{\bean}{\begin{eqnarray*}}
\newcommand{\eean}{\end{eqnarray*}}

\newcommand{\btab}[1]{\begin{tabular}{#1}}
\newcommand{\etab}{\end{tabular}}

\newcommand{\comment}[1]{}

\newcommand{\qed}{\nobreak \ifvmode \relax \else
      \ifdim\lastskip<1.5em \hskip-\lastskip
      \hskip1.5em plus0em minus0.5em \fi \nobreak
      \vrule height0.75em width0.5em depth0.25em\fi}

\def\beqa{\begin{eqnarray}}
\def\eeqa{\end{eqnarray}}

\newcolumntype{C}[1]{>{\centering\arraybackslash}m{#1}}

\def\makeatletter{\catcode`\@=11}
\makeatletter
\def\mathbox#1{\hbox{$\m@th#1$}}%
\def\math@ccstyles#1#2#3#4#5#6#7{{\leavevmode
     \setbox0\mathbox{#6#7}%
     \setbox2\mathbox{#4#5}%
     \dimen@ #3%
     \baselineskip\z@\lineskiplimit#1\lineskip\z@
     \vbox{\ialign{##\crcr
            \hfil \kern #2\box2 \hfil\crcr
            \noalign{\kern\dimen@}%
            \hfil\box0\hfil\crcr}}}}
\def\mathaccstyles{\math@ccstyles\maxdimen}
\def\maththroughstyles{\math@ccstyles{-\maxdimen}}
\def\unity%
{\maththroughstyles{.45\ht0}\z@\displaystyle {\mathchar"006C}\displaystyle 1}

\title{Non-Invertible Symmetries in Supergravity}
\author[]{Eduardo García-Valdecasas}
\affiliation[]{$^1$Jefferson Physical Laboratory, Harvard University,\\
Cambridge, MA 02138, USA}
\affiliation[]{$^2$Universidad Autonoma de Madrid, Ciudad Universitaria de Cantoblanco,\\ 28049 Madrid, Spain}

\emailAdd{egarciavaldecasastenreiro@fas.harvard.edu}

\abstract{}

\begin{document}

\abstract{Non-invertible symmetries have been extensively studied in quantum field theories in recent years. In this note we initiate their study in supergravity. We find infinite families of non-invertible defects in 11d and 10d Type II supergravities. These operators display a rich action on different probe branes. We comment on how these symmetries are removed in the UV completion, M-theory and Type II String Theory and how their existence strengthens the link between the absence of global symmetries in Quantum Gravity and the Completeness Hypothesis.  }

\maketitle


\section{Introduction}

Symmetry is one of the basic principles in contemporary physics and quantum field theory in particular. A modern definition of symmetry, pioneered in \cite{Gaiotto:2014kfa}, identifies symmetry with the topological sector of the theory. This is precisely what makes it a universal notion that helps classify quantum field theories and is robust under smooth deformations such as the RG flow. In more detail, the symmetry of a theory is given by the set of all topological operators and their fusion rules. In the most general definition one allows for topological operators of any codimension and fusion rules that may not obey a group law. The usual symmetries such as baryon number are generated by topological operators of codimension 1, $U_g (\Sigma_{d-1})$, hence acting on Hilbert space, that obey a group-law fusion $U_{g_1} (\Sigma_{d-1}) \times U_{g_2} (\Sigma_{d-1}) = U_{g_1 \cdot g_2} (\Sigma_{d-1})$. Allowing for topological operators of different codimensions gives rise to higher-form symmetries. A prototypical example is the 1-form symmetry of free Maxwell theory acting on Wilson loops. Allowing for non group-laws brings non-invertible symmetries into the game. These are generated by topological operators that need not have an inverse, hence the name. In this work we will be interested in $p$-form symmetries generated by codimension $p+1$ operators with various $p's$ and obeying non-invertible fusion rules.

While non-invertible symmetries may look exotic at first, in recent years it has been understood that they are essentially ubiquitous. In fact, theories as simple as free Maxwell in $4d$ are plagued with non-invertible symmetry operators. For a partial list of these recent developments see \cite{Rudelius:2020orz,Heidenreich:2021xpr,Nguyen:2021yld,Koide:2021zxj,Choi:2021kmx,Kaidi:2021xfk,Cordova:2022rer,Benini:2022hzx,Roumpedakis:2022aik,Bhardwaj:2022yxj,Arias-Tamargo:2022nlf,Hayashi:2022fkw,Choi:2022zal,Kaidi:2022uux,Choi:2022jqy,Cordova:2022ieu,Antinucci:2022eat,Bashmakov:2022jtl,Damia:2022rxw,Damia:2022bcd,Choi:2022rfe,Bhardwaj:2022lsg,Bartsch:2022mpm,Lin:2022xod,GarciaEtxebarria:2022vzq,Apruzzi:2022rei,Heckman:2022muc,Freed:2022qnc,Niro:2022ctq,Kaidi:2022cpf,Mekareeya:2022spm,Antinucci:2022vyk,Chen:2022cyw,Bashmakov:2022uek,Karasik:2022kkq,Cordova:2022fhg,GarciaEtxebarria:2022jky}. For a partial list of earlier results in $2d$ see \cite{Verlinde:1988sn,Petkova:2000ip,Fuchs:2002cm,Frohlich:2004ef,Bhardwaj:2017xup,Tachikawa:2017gyf,Chang:2018iay,Thorngren:2019iar,Gaiotto:2020iye,Komargodski:2020mxz,Nguyen:2021naa,Thorngren:2021yso}. While these symmetries have been extensively studied in quantum field theory, they are almost uncharted territory in supergravity\footnote{See \cite{GarciaEtxebarria:2022jky} for a noteworthy exception.}. In this work we make a first, if superficial, approach to these new lands by studying some non-invertible defects in 11d, 10d Type IIA and Type IIB supergravity. The core idea is that Chern-Simons terms, which are usually believed to imply explicit breaking of the higher form symmetries of the gauge fields \cite{Montero:2017yja}, are many times just a signal of their non-invertibility. Let us illustrate the argument, already formulated in \cite{Damia:2022bcd} by building on earlier work \cite{Cordova:2022ieu,Choi:2022jqy}. In the absence of Chern-Simons terms, gauge theories typically have equations of motion and Bianchi identities of the form,
\begin{equation}
    d \star F = 0, \quad \quad dF=0
\end{equation}
These equations imply the existence of higher-form electric and magnetic symmetries. Adding Chern-Simons terms generically spoils these symmetries, particularly the electric one, as the equation of motion now takes the form of a non-conservation equation, for instance:
\begin{equation}
    d \star F = F \wedge F
\end{equation}
However, as we discuss in more detail in the following section, in many cases the non-conservation is mild enough, since the right hand side vanishes in trivial topology, that one can still define topological operators implementing the symmetry. The price to pay is that one needs to dress the operator with some topological degrees of freedom that generate a non-invertible fusion rule. Given that supergravity theories are plagued with Chern-Simons terms\footnote{Many times these Chern-Simons terms are actually absorbed in modified Bianchi identities.} one expects a very rich set of non-invertible symmetries! These symmetries will naturally act on probe branes.

An important lesson from Quantum Gravity is that exact global symmetries are incompatible with it \cite{Banks:1988yz,Banks:2010zn,Harlow:2018tng,Harlow:2020bee,Chen:2020ojn,Hsin:2020mfa,Sasieta:2022ksu,Bah:2022uyz}. This implies that all the symmetries described in this work must be broken by the UV completion, be it M-theory or Type II String Theory. As we will see this is easily achieved by the presence of dynamical branes, in analogy to what happens with less exotic symmetries \cite{Heidenreich:2020pkc}. For further discussions on non-invertible symmetries and Quantum Gravity see \cite{Rudelius:2020orz,Heidenreich:2021xpr,Arias-Tamargo:2022nlf}.

The remainder of this note is organised as follows. In \cref{Sec:Review} we review non-invertible symmetries in quantum field theories with Chern-Simons terms. In \cref{Sec:Supergravity} we present infinite families of non-invertible defects for 11d supergravity and the 10d Type II supergravities. We study their action on probe branes in some detail. In \cref{Sec:DiffAproach} we elaborate in a different approach to the same question. We find a different set of topological operators and comment on their connection to the former ones. We conclude in \cref{Sec:Discussion} with comments on our results, their implications and an outlook of future work. We leave for the \Cref{AppTQFT,Sec:AppConstruction} the explicit construction of the TQFT's that are used in the main text to construct the general topological operators.

\section{Review of QFT examples} \label{Sec:Review}

\begin{figure}       \centering  \includegraphics[width=0.5\textwidth]{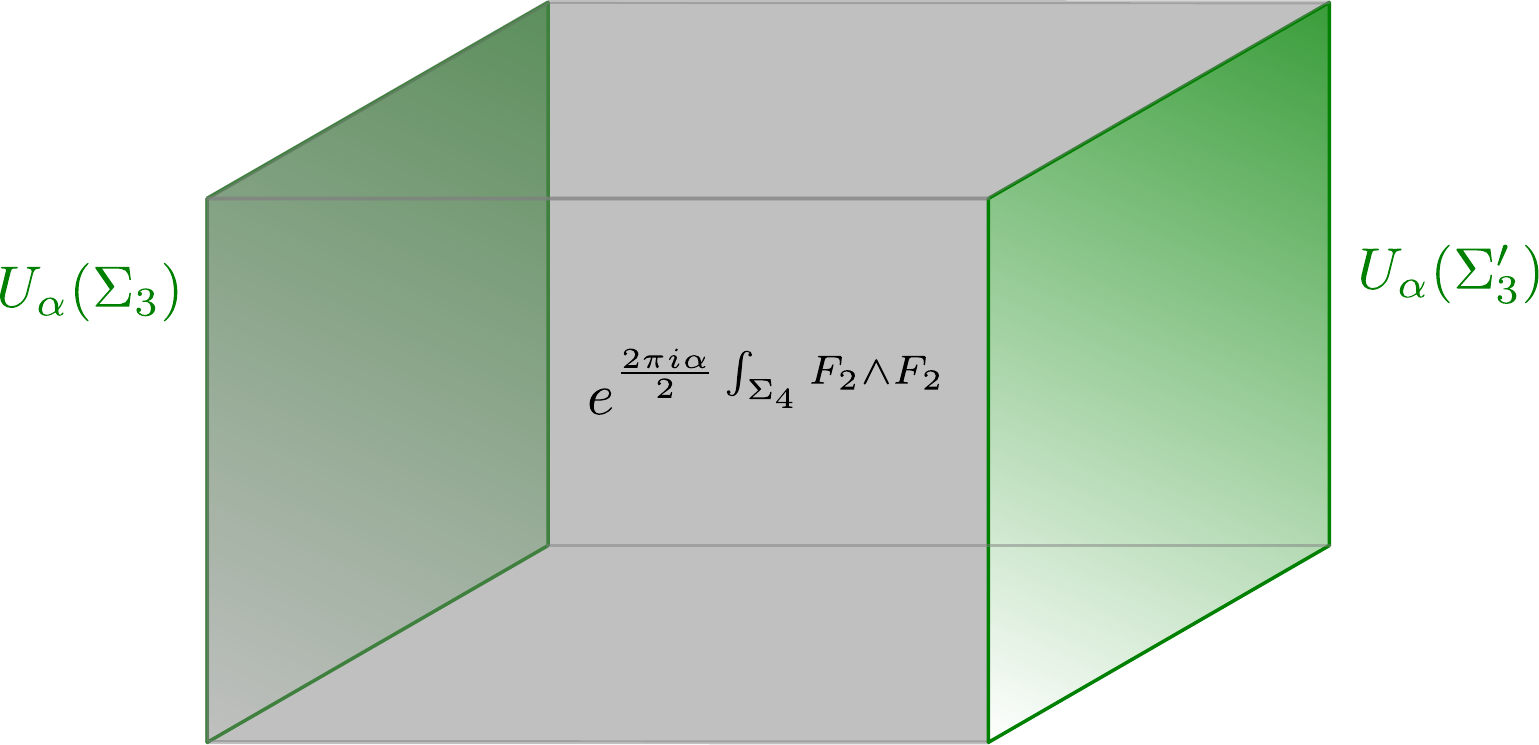} 
    \caption{A deformation of the naive operator $U_{\alpha} (\Sigma_3)$ generates an anomalous phase in the region swept by it.}  	 \label{Fig:AnomalousPhase}
\end{figure}

Chern-Simons terms typically turn the equations of motion for gauge fields into non-conservation equations for the currents they carry. Until recently, it was believed that these non-conservation equations implied the explicit breaking of the symmetry obtained by integrating the no longer conserved current. However, when the non-conservation is mild enough, one may construct topological operators for the symmetry by appropriately dressing them with gaped degrees of freedom. More concretely, consider a non-conservation equation for a $U(1)$ current $J_p$ of the form,
\begin{equation}
    d \star J_p = G_{d-p+1}
\end{equation}
where $G_{d-p+1}$ is a wedge product of gauge field strengths. If $G_{d-p+1}$ is locally exact $G_{d-p+1} = d K_{d-p}$ and its integral vanishes in manifolds of trivial topology\footnote{With trivial topology we refer to manifolds homeomorphic to $\mathbb{R}^4$ or $S^4$.}, one can safely improve the current,
\begin{equation}
    d (\star J_p - K_{d-p}) = 0
\end{equation}
This improved current is conserved but may be non gauge invariant. The integrated charge for such a current is known as a Page charge \cite{Marolf:2000cb} and we can use it to write an operator,
\begin{equation}
    e^{2 \pi i \alpha \int_{\Sigma_{d-p}} \star J_p - K_{d-p}   }
\end{equation}
For simple enough spacetime manifolds (trivial topology) this operator is actually topological and well defined. In fact, in most cases\footnote{An exception is the case of $G_{d-p+1}$ being a single field strength, as in 3d Chern-Simons theory or 4d BF theory. For these theories however, $\int_{M_2} G_{d-p+1} = \int_{M_2} F \neq 0$, so our assumptions are not satisfied.} a new operator can be introduced with $K_{d-p}$ modified to be also topological and gauge invariant in arbitrary manifolds. Let us consider a particularly simple example, 5d Chern-Simons Maxwell theory \cite{Damia:2022bcd},
\begin{equation}
    S = 2 \pi \int -\frac{1}{2} F_2 \wedge \star F_2 + \frac{1}{6} A_1 \wedge F_2 \wedge F_2
\end{equation}
Note that we have chosen to stick to the conventions in the supergravity literature. In particular, we have chosen field strengths to have integer periods, $\int F \in \mathbb{Z}$. In the absence of a Chern-Simons term, this theory would have $U(1)^{(1)}_e \times U(1)_m^{(2)}$ electric and magnetic symmetries with currents $ j_e = F_2, \, j_m = \star F_2$. These currents are conserved due to the equation of motion and the Bianchi identity, respectively. However, the Chern-Simons term seems to break the electric symmetry: the equation of motion becomes,
\begin{equation}
    d \star F_2 =  \frac{1}{2} F_2 \wedge F_2
\end{equation}
This equation implies that $F_2$ is not conserved anymore, so the naive operator,
\begin{equation}
     U_{\alpha} (\Sigma_3) = \text{exp} \left( 2 \pi i \alpha \int_{\Sigma_3} \star F_2     \right) \label{Eq:BadOpeNaive}
\end{equation}
is no longer topological. However, $F_2 \wedge F_2$ is locally exact and, since there is no $U(1)$ instanton number in $S^4$, its integral vanishes in manifolds of trivial topology\footnote{This statement is sensitive to the UV completion of the quantum field theory at hand. For instance, there are stringy instantons in single D-branes in String Theory \cite{Petersson:2007sc} and there are also $U(1)$ instantons in non-commutative spacetimes \cite{Seiberg:1999vs}. We thank Iñaki García-Etxebarria for raising this point. }. We may then improve the current,
\begin{equation}
    d \left(  \star F_2 -\frac{1}{2} A_1 \wedge F_2 \right) = 0
\end{equation}
which is conserved but not gauge invariant. The corresponding topological operator, only valid for simple enough topology, is,
\begin{equation}
    \Tilde{U}_{\alpha} (\Sigma_3) = \text{exp} \left( 2 \pi i \alpha \int_{\Sigma_3} \star F_2 - \frac{1}{2} A_1 \wedge F_2       \right) \label{Eq:BadOpe}
\end{equation}
While this operator is not valid for arbitrary topology, if $\alpha \in [0,1)$ is rational, one can define an improved operator which is topological and gauge invariant for arbitrary $\Sigma_3$. For the particular case of $\alpha = 1/N$, with $N \in \mathbb{Z}$, the corresponding operator is,
\begin{equation}
    \mathcal{D}_{1/N}(\Sigma_3) = \int D c_1 \Big|_{\Sigma_3} \text{exp} \left( 2 \pi i \oint \frac{\star F_2}{N} + \frac{N}{2} c_1 \wedge d c_1 -  c_1 \wedge F_2  \right) \label{Eq:TopOpCS}
\end{equation}
where $c_1$ is a $U(1)$ gauge field localized in the topological defect. That this operator is the gauge invariant version of \ref{Eq:BadOpe} can be morally seen by integrating out $c_1 = A_1/N$. While this integration is not a correct equation for $A_1,c_1$ which are both $U(1)$ gauge fields, it gives the correct result. For more details, including an explicit construction of the defect using higher half-space gauging see \cite{Damia:2022bcd}. For a general $\alpha = p/N$, $p,N \in \mathbb{Z}$, a good topological operator can also be written in terms of the minimal $\mathcal{A}^{n,p}$ TQFT as in \Cref{Eq:GeneralOp}. For more details on $\mathcal{A}^{n,p}$, see \cite{Hsin:2018vcg}. What we have done is to dress the naive operator in \ref{Eq:BadOpeNaive} in such a way that the new degrees of freedom cancel the anomalous phase it generates as it is deformed,
\begin{equation}
    U_{\alpha} (\Sigma^{\prime}_3) = U_{\alpha} (\Sigma_3) e^{\frac{2 \pi i \alpha}{2} \int_{\Sigma_4} F_2 \wedge F_2}
\end{equation}
For a pictorial representation see \Cref{Fig:AnomalousPhase}. 
The upshot is that the rational $\alpha = p/N$ subgroup of the $U(1)$ electric 1-form symmetry survives and is generated by \ref{Eq:GeneralOp}. The price to pay is that the symmetry becomes non-invertible, as can be seen by explicitly computing the fusion rules of the topological operators. These operators act on Wilson lines as the naive operators would have, but they also act on 't Hooft surfaces by attaching a fractional flux to them. The action is completely analogous to what we will describe for 11d supergravity in \cref{Sec:Mtheory}. We will denote these non-invertible symmetries with rational valued parameters as $\Gamma^{(1)}_{\mathbb{Q}}$. 
The above construction generalizes to mixed Chern-Simons terms \cite{Damia:2022bcd}. Consider for instance,
\begin{equation}
    S = 2 \pi \int -\frac{1}{2} F_2 \wedge \star F_2 - \frac{1}{2} G_2 \wedge \star G_2 + \frac{1}{2} C_1 \wedge F_2 \wedge F_2
\end{equation}
Where $F_2= dA_1$, $G_2 = dC_1$. In the absence of the Chern-Simons coupling, this theory has electric and magnetic symmetries 
\begin{equation}
    U(1)^{(1)}_{e,A} \times U(1)_{m,A}^{(2)}\times U(1)^{(1)}_{e,C} \times U(1)_{m,C}^{(2)}
\end{equation}
Yet again, the Chern-Simons coupling spoils the conservation equations of the electric symmetries, which become,
\begin{equation}
    d \star F_2 = G_2 \wedge F_2, \quad \quad d \star G_2 =\frac{1}{2} F_2 \wedge F_2
\end{equation}
In the same vein as before, these two currents can be improved to Page currents. In turn, topological operators corresponding to the Page charge can be appropriately defined. For $\alpha = 1/N$ these operators can be written as,
\begin{align}
    \mathcal{D}^G_{1/N}(\Sigma_3) = \int D c_1 \Big|_{\Sigma_3} \text{exp} \left( 2 \pi i \oint \frac{\star G_2}{N} + \frac{N}{2} c_1 \wedge d c_1 -  c_1 \wedge F_2  \right) \label{Eq:TopoOp35d}\\
    \mathcal{D}^F_{1/N}(\Sigma_3) = \int D c_1 D v_1 \Big|_{\Sigma_3} \text{exp} \left(2 \pi i \oint \star \frac{F_2}{N} + N v_1 \wedge dc_1 -  c_1 \wedge G_2  -  v_1 \wedge F_2  \right) \label{Eq:TopoOp25d}
\end{align}
These two constructions can once again be extended to any rational $\alpha = p/N$ by using the $\mathcal{A}^{n,p}$ theories. The upshot is  that the electric symmetries are turned non-invertible by the Chern-Simons terms and the true symmetry of the theory is,
\begin{equation}
  \Gamma^{(1)}_{\mathbb{Q},A}  \times U(1)_{m,A}^{(2)}\times\Gamma^{(1)}_{\mathbb{Q},C} \times U(1)_{m,C}^{(2)}
\end{equation}
In the remainder of this note we explore similar constructions in supergravity, where Chern-Simons terms are ubiquitous.

\section{Non-invertible symmetries in Supergravity} \label{Sec:Supergravity}
\subsection{11d Supergravity} \label{Sec:Mtheory}

The construction of the non-invertible defects in the previous examples is only possible thanks to the existence of a particular 3d TQFT with the appropriate 1-form symmetry and anomaly. For would-be $U(1)$ actions with phase $\alpha = 2\pi/N$ the full topological operator is \ref{Eq:TopoOp35d} and the TQFT that is stacked on top of the naive operator \ref{Eq:BadOpeNaive} takes the form,
\begin{equation}
    \mathcal{A}^{N,1}_3 (\Sigma_3) =  \int  D c_1 \Big|_{\Sigma_3} \text{exp} \left( 2 \pi i \oint_{\Sigma_3} \frac{N}{2} c_1 \wedge dc_1 -  c_1 \wedge F_2  \right)
\end{equation}
where $c_1$ is an auxiliary $U(1)$ gauge field living in $\Sigma_3$ and $F_2$ is the magnetic current in the bulk that needs to be gauged to obtain the defect. This TQFT, and its generalizations $\mathcal{A}^{N,p}_3 (\Sigma_3)$, are called fractional quantum Hall states, or FQHE states, and are particular to 3d. In 7d an analog FQHE construction con be made\footnote{See, for instance \cite{Heckman:2017uxe}.}, where one can write the following TQFT,
\begin{equation}
   \mathcal{A}^{N,1}_7(\Sigma_7) =  \int  D c_3 \Big|_{\Sigma_7} \text{exp} \left( 2 \pi i \oint_{\Sigma_7} \frac{N}{2} c_3 \wedge dc_3 -  c_3 \wedge F_4  \right)
\end{equation}
As we now argue, this TQFT precisely arises in 11d supergravity. Consider its action,
\begin{equation}
S = \frac{1}{2 \kappa^2_{11}} \int_{\Sigma_{11}} \sqrt{-g} R - \frac{1}{2} F_4 \wedge \star F_4 - \frac{1}{6} A_3 \wedge F_4 \wedge F_4
\end{equation}
Where $2 \kappa_{11}^2 = (2\pi)^{-1} (2 \pi l_p)^9$. If the Chern-Simons coupling is turned off, this theory has 2 symmetries $U(1)_e^{(3)} \times U(1)_m^{(6)}$ with currents $  j_e = F_4, \, j_m = \star F_4$ conserved thanks to the equation of motion and the Bianchi identity of $F_4$. As discussed in \cite{Heidenreich:2020pkc} it also has a Chern-Weil symmetry with current $F_4 \wedge F_4$, which will not be relevant for our purposes. Once one includes the CS interaction the conservation equation of $U(1)_e^{(3)}$ is modified to,
\begin{equation}
    d \star F_4 = \frac{1}{2} F_4 \wedge F_4 \label{Eq:EOMMtheory}
\end{equation}
Hence the CS term explicitly breaks $U(1)_e^{(3)}$ and gauges the Chern-Weil current. By now the game we must play is clear, this current fulfills the conditions to be improved to a $U(1)$ Page current,
\begin{equation}
    d \left (\star F_4 - \frac{1}{2} A_3 \wedge F_4 \right) = 0 
\end{equation}
which is conserved but not gauge invariant. The naive operator, which is only valid for trivial topology is,
\begin{equation}
    U_{\alpha} (\Sigma_7) = \text{exp} \left( 2 \pi i \alpha \int_{\Sigma_7} \star_{11} F_4 - \frac{1}{2} A_3 \wedge F_4       \right) \label{Eq:BadOpeM}
\end{equation}
Writing the corresponding good topological operator is straightforward. For $\alpha = 1/N \in U(1)$ we just need to stack the 7d FQHE theory. The resulting operator is,
\begin{equation}
    \mathcal{D}_{1/N}(\Sigma_7) =  \int  D c_3 \Big|_{\Sigma_7} \text{exp} \left( 2 \pi i \oint_{\Sigma_7} \frac{\star_{11} F_4}{N} +\frac{N}{2} c_3 \wedge dc_3 -  c_3 \wedge F_4  \right)
\end{equation}
In fact, we can make use of the $\mathcal{A}^{(N,p)}_7[b_4]$ theory defined in \Cref{AppTQFT} to build a topological operator for any $\alpha \in [0,1)$ of the form $\alpha = p/N$, with $p,N \in \mathbb{Z}$,
\begin{equation}
    \mathcal{D}_{p/N}(\Sigma_7) = \text{exp} \left( \frac{2 \pi i p}{N}  \int_{\Sigma_7} \star_{11} F_4 \right)\times \mathcal{A}^{(N,p)}_7 \left[\frac{F_4}{N} \right] \label{Eq:TopoDefectMtheory}
\end{equation}
These defects can be explicitly constructed by higher gauging the magnetic $U(1)_m^{(6)}$ symmetry, as detailed in \cref{Sec:AppConstruction}. The upshot of the discussion above is that, contrary to expectations, the electric $U(1)_e^{(3)}$ symmetry is not completely broken, but a non-invertible $rational$ discrete subgroup remains. The symmetries associated to $F_4$ in M-theory are then,
\begin{equation}
    \Gamma_{\mathbb{Q}}^{(3)} \times U(1)_m^{(6)}
\end{equation}
Of course we expect these two symmetries to be broken by the UV-completion of 11d supergravity. This is indeed the case in M-theory, as inclusion of dynamical M2- and M5-branes breaks them explicitly. In fact, given that one needs to gauge the magnetic symmetry to build the electric defects, the presence of dynamical M5-branes is enough to break both symmetries. A consequence is that $\Gamma_{\mathbb{Q}}^{(3)}$ must be broken at an energy scale lower or equal than $U(1)_m^{(6)}$.

\begin{figure}       \centering  \includegraphics[width=0.6\textwidth]{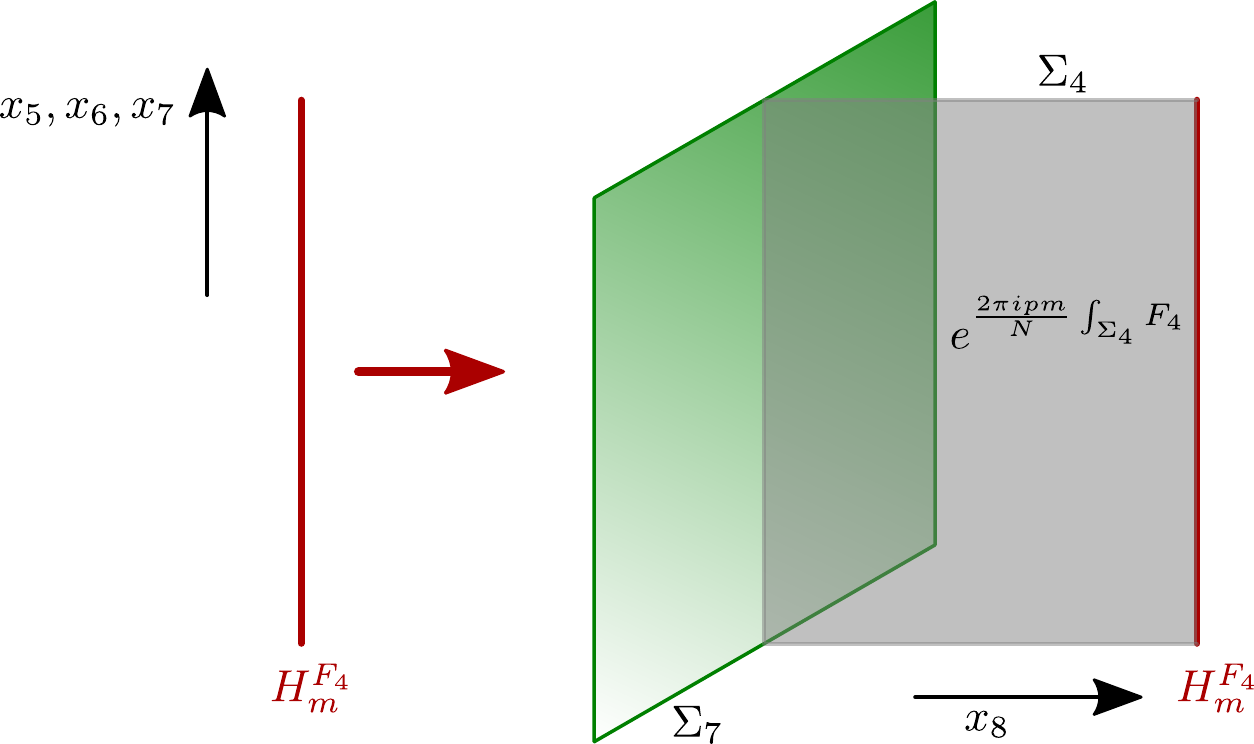} 
    \caption{A magnetic source for $F_4$ crosses the topological defect and picks up a fractional $F_4$ flux attached to it. }  	 \label{Fig:M-theoryDefect}
\end{figure}

Let us now study more carefully the action of the topological defect \ref{Eq:TopoDefectMtheory} on electric and magnetic probes, which we call probe M2- and M5-branes for obvious reasons. On probe M2-branes, which source $\star F_4$, the action is invertible and given by the first term in \ref{Eq:TopoDefectMtheory}. The action is indistinguishable from the one we expected for the electric 3-form symmetry. The second term in \ref{Eq:TopoDefectMtheory} can detect magnetic charges, sources for $F_4$. In order to find the precise action consider the construction of the topological defect in $\Sigma_7$ via higher gauging of the magnetic symmetry as explained in \Cref{Sec:AppConstruction}. Take 11d spacetime to be $\mathbb{R}^{11}$ with coordinates $x_1, ..., x_{11}$ such that $\Sigma_7$ spans $x_1, ... , x_7$ and the gauging is defined for $x_8 > 0$ such that $\Sigma_8 = \mathbb{R}^7 \times \mathbb{R}^{>0}_{x_8}$. Denote by $H_m^{F_4}$ a 6-dimensional source of $m$ units of $F_4$ flux, which could be a bunch of probe M5-branes, spanning directions $x_{5,6,7,9,10,11}$. If we displace $H_m^{F_4}$ from $x_8 < 0$ to $x_8 > 0$ it enters into the region where the magnetic symmetry is gauged and it stops being gauge invariant. In more detail, the 3-dimensional part of the probe M5 worldvolume that is inside the submanifold where the symmetry is gauged, $\Sigma_3 \equiv \mathbb{R}^3_{x_5, x_6, x_7} = WV(M5) \cap \Sigma_8$,  transforms under $b_4 \rightarrow b_4 + d \Lambda_3$ gauge transformations by picking up a phase,
\begin{equation}
    H_m^{F_4} \rightarrow H_m^{F_4} e^{2 \pi i m \int_{\Sigma_3} \Lambda_3} , \quad \quad \text{for} \quad x_8 > 0
\end{equation}
Hence, in the $x_8 > 0$ region the gauge invariant object is,
\begin{equation}
    H_m^{F_4}  e^{ - 2 \pi i m \int_{\Sigma_4} b_4} =  H_m^{F_4}  e^{\frac{2 \pi ipm}{N} \int_{\Sigma_4} F_4}
\end{equation}
where $\Sigma_4$ is such that $\partial \Sigma_4 = \Sigma_3$. To write the right hand side we have used the  equation of motion for $b_4$ in \ref{Sec:AppConstruction} $mod \, \, N$. We conclude that the defect acts on the magnetic source by attaching a fractional $F_4$ flux along $\Sigma_4$, as depicted in Fig. \ref{Fig:M-theoryDefect},
\begin{equation}
    \mathcal{D}_{p/N}(\Sigma_7)  H_m^{F_4} = H_m^{F_4}  e^{\frac{2 \pi ipm}{N} \int_{\Sigma_4} F_4}.
\end{equation}
An important consequence of this action is that, if $m \neq 0 \, \text{mod} \, N$, the symmetry defect annihilates the magnetic source, as argued in \cite{Choi:2022fgx}. Consider again \Cref{Fig:M-theoryDefect}. The idea is that, since $\mathcal{D}_{p/N}(\Sigma_7)$ is topological, it may be shrunk to a point and removed, giving rise to a topological endpoint for $V(\Sigma_4)_{\frac{pm}{N}} \equiv e^{\frac{2 \pi ipm}{N} \int_{\Sigma_4} F_4}$. This endpoint is local and its existence implies that $V(\Sigma_4)_{\frac{pm}{N}}$ can develop a hole and disappear. However, $V(\Sigma_4)_{\frac{pm}{N}}$ is  a generator of the magnetic symmetry $U(1)_m^{(6)}$ which acts faithfully, so it better be that it can't just go away! The only solution is that correlation functions with this endpoint give zero. We conclude that the topological operator annihilates $H_m^{F_4}$ sources unless, 
\begin{equation}
    \frac{pm}{N} \in \mathbb{Z} 
\end{equation}
However, if $m = 0 \, \text{mod} \, N$, $V(\Sigma_4)_{\frac{pm}{N}}$ becomes trivial except at the boundary of $\Sigma_4$ and the action of the topological operator leaves an operator insertion in $\Sigma_3$. We leave the study of this junction in detail for future work.

\subsection{Type IIA Supergravity}
The existence of non-invertible symmetries in 11d supergravity suggests the existence of appropriate counterparts in Type IIA supergravity, which arises when dimensionally reducing on a circle. We will not attempt direct dimensional reduction of the symmetries in this work. We study instead the non-invertible symmetries directly from the 10 formulation. Consider the low energy action of type IIA string theory in 10d \cite{Becker:2006dvp},
\begin{align}
\begin{split}
    S_{IIA} = \frac{1}{2 \kappa^2}\int_{M^{10}} \sqrt{-g} \left[ e^{-2\Phi} \left( R + 4|d \Phi|^2 - \frac{1}{2} |H_3|^2 \right) -\frac{1}{2} |F_2|^2 -\frac{1}{2} |\Tilde{F}_4|^2 \right] - \\ - \frac{1}{2 \kappa^2}\int_{M^{10}} \frac{1}{2} B_2 \wedge F_4 \wedge F_4
\end{split}
\end{align}
where $\Tilde{F}_4 = dA_3 + A_1 \wedge H_3$ is invariant under the gauge transformation $(A_1, A_3) \rightarrow (A_1 + d\lambda_0, A_3 -\lambda_0 \wedge  H_3)$. The equations of motion for $F_2, \Tilde{F}_4$ and $H_3$ are,
\begin{align}
    d \star F_2 &= H_3 \wedge \star \Tilde{F}_4 \label{Eq:EOMF2} \\
    d \star \Tilde{F}_4 &=  H_3 \wedge F_4 =  H_3 \wedge \Tilde{F}_4 \label{Eq:EOMF4}\\
        d \star H_3 &= \frac{1}{2} \Tilde{F_4} \wedge \Tilde{F_4} + F_2 \wedge \star \Tilde{F}_4 \label{Eq:EOMH3}
\end{align}
The Bianchi Identities are, 
\begin{equation}
    d H_3 = 0, \quad      d F_2 = 0, \quad d \Tilde{F}_4 =  F_2 \wedge H_3
\end{equation}
The Bianchi identities for $F_2, H_3$ imply that there are two conserved currents: $\star H_3$, $\star F_2$. These generate two magnetic symmetries $U(1)^{(6)}_m \times U(1)^{(7)}_m$ which are well understood, see for instance \cite{Heidenreich:2020pkc}.  The remaining equations can be rewritten by introducing Hodge dual field strengths $ \tilde{F}_p = \star \Tilde{F}_{10-p}$ as,
\begin{align}
    d  \Tilde{F}_4 =  F_2 \wedge H_3, \quad d  \Tilde{F}_6 =  \Tilde{F}_4 \wedge H_3 , \quad d  \Tilde{F}_8 =  \Tilde{F}_6 \wedge H_3 , \label{Eq:EOMTypeIIA} \\
    d \star H_3 =  \frac{1}{2} \Tilde{F}_4 \wedge \Tilde{F}_4 + F_2 \wedge  \Tilde{F}_6 \label{Eq:EOMTypeIIAH3}
\end{align}
The newly introduced gauge invariant field strengths are defined as $\Tilde{F}_p = d A_{p-1} + A_{p-3} \wedge H_3$. From the equations above we see that there are three candidates for non-invertible symmetries. Indeed, the right hand side in equations \ref{Eq:EOMTypeIIA} can be shown to be locally exact and its integral to vanish in manifolds with trivial topology, so we may write the following conserved but gauge variant Page currents,
\begin{align}
    d \left( \Tilde{F}_4  - A_1 \wedge H_3 \right)    = 0 \label{Eq:ImprovedF4} \\
    d \left( \Tilde{F}_6  - A_3 \wedge H_3 \right) = 0 \label{Eq:ImprovedF6}\\
    d \left( \Tilde{F}_8  - A_5 \wedge H_3 \right) = 0 \label{Eq:ImprovedF8}
\end{align}
At this point the strategy seems clear: we can dress the would-be topological operators with some gaped degrees of freedom to obtain a good topological operator. However, for \cref{Eq:ImprovedF6,Eq:ImprovedF8} things are not so simple. Let us focus first in \cref{Eq:ImprovedF4}. We consider a U(1) transformation with $\alpha = 1/N$ and in analogy to \ref{Eq:TopoOp25d} propose the following operator,
\begin{align}
    \mathcal{D}^{F_4}_{1/N}(\Sigma_4) = \int D c_1 D v_2 \Big|_{\Sigma_4} \text{exp} \left(2 \pi i \oint_{\Sigma_4}  \frac{\tilde{F}_4}{N} + N v_2 \wedge dc_1 -  c_1 \wedge H_3 -  v_2 \wedge dA_1  \right) \label{Eq:TopOpIIA1}
\end{align}
This operator is gauge invariant and can be built explicitly by gauging a $\mathbb{Z}^{(6)}_N \times \mathbb{Z}^{(7)}_N$ subgroup of the magnetic symmetry $U(1)^{(6)}_m \times U(1)^{(7)}_m$ with currents $\star H_3 $ and $\star F_2$ in a manifold $\Sigma_5$ such that $\partial \Sigma_5 = \Sigma_4$, as shown in \Cref{Sec:AppConstruction}. In fact, as discussed there, the operator can be built for arbitrary $\alpha = p/N$ in a similar way, such that it is explicitly topological and gauge invariant. We hence conclude that the would-be magnetic $U(1)^{(5)}_m$ symmetry of $\Tilde{F}_4$, which is broken by the modified Bianchi identity, becomes a non-invertible $\Gamma_{\mathbb{Q}}^{(5)}$ symmetry. Let us describe how the topological operator acts on the different probe branes. Since this discussion is a straightforward extension of the 11d supergravity we will be brief. A related in-depth discussion for the case of axion electrodynamics is presented in \cite{Choi:2022jqy,Yokokura:2022alv}. The defect acts invertibly in probe D4-branes, as the naive magnetic symmetry of $\Tilde{F}_4$ would have, but it also acts non-invertibly in probe NS5- and D6-branes. In particular, these probes are not gauge invariant if they intersect the auxiliary higher-gauging submanifold and a flux needs to be attached to them. The upshot is that the topological defect with $\alpha = p/N$ annihilates the probe NS5- or D6-branes if their charge $m$ does not satisfy the following relation,
\begin{equation}
    \frac{pm}{N} \in \mathbb{Z} \label{Eq:Condition}
\end{equation}
If the relation above is satisfied, the action of the topological defect on the probe NS5- or D6-brane leaves behind an operator stuck in their worldvolume of dimension $1$ for the NS5-brane and $2$ for the D6-brane. While the study of these junctions is very interesting, it goes beyond the scope of this work and we leave it for the future. We expect that anomaly inflow from the bulk will impose the existence of certain worldvolume degrees of freedom on the branes that will in turn be used to furnish the appropriate operators.

Consider now the Page currents in \cref{Eq:ImprovedF6,Eq:ImprovedF8}. As we now explain, the would-be topological operator analogous to \ref{Eq:TopOpIIA1} can't be built by higher gauging because there is a higher anomaly. For \ref{Eq:ImprovedF6}, for instance, one could propose the following operator,
\begin{align}
    \mathcal{D}^{F_6}_{1/N}(\Sigma_6) = \int D c_3 D v_2 \Big|_{\Sigma_6} \text{exp} \left(2 \pi i \oint_{\Sigma_6}  \frac{\tilde{F}_6}{N} + N v_2 \wedge dc_1 -  c_3 \wedge H_3 -  v_2 \wedge dA_3  \right) \label{Eq:TopOpIIA2}
\end{align}
One immediately notices however, that the last term is not invariant under $(A_1,A_3) \rightarrow (A_1 + d \lambda_0, A_3 - \lambda_0 \wedge H_3)$ gauge transformations. One can gain further insight into this issue by explicitly introducing $\mathbb{Z}_N$ background gauge fields for $\star H_3$ and $\star d A_3$ in $\Sigma_7$ such that $\partial \Sigma_7 =  \Sigma_6$. If correct, this higher gauging should generate \ref{Eq:TopOpIIA2}. There is, at first sight, nothing wrong in gauging the currents $\star H_3$ and $\star dA_3$ since they are both conserved, the gauging is implemented by adding the following terms to the action\footnote{For more details on this procedure see \cref{Sec:AppConstruction}.},
\begin{equation}
    \delta S = 2 \pi i \int_{\Sigma_7} N p^{\prime} b_3 \wedge b_4 + N b_3 \wedge d \hat{c}_3 + N b_4 \wedge d \hat{v}_2  - b_4 \wedge H_3 - b_3 \wedge d A_3
\end{equation}
The problem is that under an $A_1$ gauge transformation  the last term picks up an anomalous piece,
\begin{equation}
    2 \pi i \int_{\Sigma_7} b_3 \wedge d \lambda_0 \wedge H_3 
\end{equation}
We conclude that there is a higher anomaly\footnote{This anomaly is in fact inherited from a standard anomaly in the bulk theory given by inflow as $\sim B_6 \wedge d A_1 \wedge H_3$, $B_6$ being the background coupled to the would-be current $\star dA_3$.} encoded by inflow as $2 \pi i \int b_3 \wedge d A_1 \wedge H_3 $ which precludes the higher gauging of the symmetry associated with the current $\star d A_3$. A similar statement holds for $\star d A_5$. A different way of phrasing the problem is by noticing that there is no magnetic symmetry associated to the would-be currents $\star dA_3$ or $\star dA_5$. We do not have the necessary symmetries for the higher gauging procedure. An open possibility is to gauge the 5-form non-invertible magnetic symmetry for $\Tilde{F}_4$ that we just built. This would amount to performing an explicit sum over the non-invertible defects in $\Sigma_7$ instead of the coupling to the background field $b_3$. We would in turn be able to do it for $\Tilde{F}_8$ by using the newly found operators. It is not clear however how to perform this gauging, which we leave for future work. In section \ref{Sec:DiffAproach} we comment on how this problem seems to be avoided in the alternative approach presented in \cite{Karasik:2022kkq,GarciaEtxebarria:2022jky}.

\subsection{Type IIB Supergravity}
The discussion above translates almost verbatim to Type IIB supergravity, which has the following action,
\begin{align}
\begin{split}
    S_{IIB} = \frac{1}{2 \kappa^2}\int_{M^{10}} \sqrt{-g} \left[ e^{-2\Phi} \left( R + 4|d \Phi|^2 - \frac{1}{2} |H_3|^2 \right) -\frac{1}{2} |F_1|^2 -\frac{1}{2} |\Tilde{F}_3|^2 -\frac{1}{2} |\Tilde{F}_5|^2 \right] - \\ - \frac{1}{4 \kappa^2}\int_{M^{10}} A_4 \wedge H_3 \wedge F_3
\end{split}
\end{align}
where gauge invariant field strengths $\Tilde{F}_3 = F_3 - A_0 \wedge H_3$ and $\Tilde{F}_5 = F_5 -\frac{1}{2} A_2 \wedge H_3 + \frac{1}{2} B_2 \wedge F_3$ have been introduced. This action needs to be supplemented with the self-duality condition for $\Tilde{F}_5$: $\Tilde{F}_5 = \star \Tilde{F}_5$. The equations of motion are:
\begin{align}
    d \star H_3 &= - \Tilde{F}_5 \wedge \Tilde{F}_3 + F_1 \wedge \star \Tilde{F}_3 \label{Eq:EOMH3-2} \\
    d \star F_1 &= -H_3 \wedge \star \Tilde{F}_3 \label{Eq:EOMF1}\\
    d \star \Tilde{F}_3 &= \Tilde{F}_5 \wedge H_3 \label{Eq:EOMF3} \\
    d \star \Tilde{F}_5 &= -\Tilde{F}_3 \wedge H_3  \label{Eq:EOMF5}
\end{align}
And the Bianchi Identities are,
\begin{align}
    d H_3 &= 0 \label{Eq:BianchiH3-2} \\
    d F_1 &= 0 \label{Eq:BianchiF1} \\
    d \Tilde{F}_3 &= - F_1 \wedge H_3 \label{Eq:BianchiF3} \\
    d  \Tilde{F}_5 &= - \Tilde{F}_3 \wedge H_3 \label{Eq:BianchiF5}
\end{align}
Note that \ref{Eq:EOMF5} and \ref{Eq:BianchiF5} are the same equation. There are two conserved currents, $\star H_3$ and $\star F_1$, which generate a $U(1)^{(6)}_m \times U(1)^{(8)}_m$ magnetic symmetry.   Armed with the experience from the previous section we recognise that only the modified Bianchi identity $\Tilde{F}_3 = -F_1 \wedge H_3$ can give rise to a non-invertible symmetry\footnote{Similar comments as in previous section apply here. One may wonder whether other modified Bianchi identities can give rise to non-invertible symmetries upon gauging the other non-invertible symmetries.}. The corresponding topological operator is,
\begin{align}
    \mathcal{D}^{F_3}_{1/N}(\Sigma_3) = \int D c_0 D v_2 \Big|_{\Sigma_3} \text{exp} \left(2 \pi i \oint_{\Sigma_3}  \frac{\tilde{F}_3}{N} - N v_2 \wedge dc_0 +  c_0 \wedge H_3 +  v_2 \wedge dA_0  \right) \label{Eq:TopOpIIB1}
\end{align}
which can be checked to correspond to the integral of the corresponding Page current upon naive integration of $c_0 = A_0/N$ and $v_2 = -B_2/N$. Once again we conclude that the $U(1)^{(5)}_m$ with non-conserved current $\star \Tilde{F}_3$ is not completely broken, but a non-invertible $\Gamma_{\mathbb{Q}}^{(5)}$ remains. The discussion regarding the action of this operator on the different probe objects in the theory is completely analogous as the one for Type IIA and we choose to free the reader from it. Let us just mention that it acts invertibly in D5-branes and non-invertibly in NS5- and D7-branes. 

\section{A different approach} \label{Sec:DiffAproach}
We have so far seen how to explicitly build non-invertible topological defects in 11d and 10d supergravity. In each of those cases, a would-be broken $U(1)$ symmetry is still realized by more exotic topological operators. The price to pay is that the topological operators only exist for rational angles $\alpha = p/N$ and their fusion rules become non-invertible. As we have discussed, the existence of the topological operators is intimately tied to the existence of precise higher form symmetries, a discrete subgroup of which can be gauged in a particular way. In this section we compare our approach to the one pioneered in \cite{Karasik:2022kkq,GarciaEtxebarria:2022jky} and applied to supergravity in \cite{GarciaEtxebarria:2022jky}. The idea is elegant and simple, consider for instance the Page charge associated to \ref{Eq:ImprovedF4},
\begin{equation}
    U_\alpha(\Sigma_4) = \text{exp} \left( 2 \pi i \alpha \int_{\Sigma_4} \Tilde{F}_4 - A_1 \wedge H_3   \right) \label{Eq:PageCharge}
\end{equation}
This current is topological but not gauge invariant under $A_1 \rightarrow A_1 + d \lambda_0$. In the preceding sections we saw that a gauge invariant version of \ref{Eq:PageCharge} can be written for $\alpha = p/N$ by stacking an appropriate TQFT. A simpler approach is to introduce a Stueckelberg-like field that restores gauge invariance. A compact scalar $\theta$ transforming under $A_1$ gauge transformations as $\theta \rightarrow \theta + \lambda_0$ does the job,
\begin{equation}
    U_\alpha^{\prime}(\Sigma_4) = \int D \theta \Big|_{\Sigma_4} \text{exp} \left( 2 \pi i \alpha \int_{\Sigma_4} \Tilde{F}_4 - (A_1 - d \theta) \wedge H_3   \right) \label{Eq:PageChargeGood}
\end{equation}
An immediate advantage of this approach is that $\alpha$ is not limited to be rational and the whole $U(1)$ symmetry is unbroken. Still this $U(1)$ can be seen to act non-invertibly on sources of $H_3$ flux. Consider the following setup, $\Sigma_4 = S^3 \times S^1$ with $m$ units of $H_3$ flux $\int_{S^3} H_3 = m$, which could be sourced by $m$ NS5-branes, for instance. We wish to evaluate the path integral with the insertion of the topological operator,
\begin{equation}
    \langle U_\alpha^{\prime} (S^3 \times S^1) \rangle = \left\langle   \int D \theta \Big|_{\Sigma_4} \text{exp} \left( 2 \pi i \alpha \int_{\Sigma_4} \Tilde{F}_4 - (A_1 - d \theta) \wedge H_3   \right)    \right\rangle
\end{equation}
We can evaluate explicitly the term $\sim d \theta H_3$ by taking $H_3$ to be a background:
\begin{equation}
    \int \mathcal{D} \theta \, \text{exp} \left( 2 \pi i \alpha \int_{S^3 \times S^1} d \theta \wedge H_3 \right) = \sum_{\omega \in \mathbb{z}} e^{2 \pi i \alpha m \omega} 
\end{equation}
where we have traded the integral over $\theta$ by a sum over its periods in $S^1$. The resulting sum is a delta function that is non-zero only for 
\begin{equation}
    \alpha m \in \mathbb{Z}
\end{equation}
An important remark is that this operator does not act on objects that only source $dA_1$. Indeed, if we consider a similar setup as before but with $\Sigma_4 = S^2 \times S^1 \times S^1$ and $m$ units of $dA_1$ flux in $S^2$, we find $\langle U_\alpha^{\prime} (S^2 \times S^1\times S^1) \rangle = 1$. This is in sharp contrast with the topological operator that we built previously, \ref{Eq:TopOpIIA1}, which acts non-invertibly both on sources of $H_3$ and $dA_1$. In order to better understand the connection between the two approaches, let us define yet another operator,
\begin{equation}
    \hat{U}_\alpha(\Sigma_4) = \int D \theta D c_1 \Big|_{\Sigma_4} \text{exp} \left( 2 \pi i \alpha \int_{\Sigma_4} \Tilde{F}_4 - \frac{1}{2} (A_1 - d \theta) \wedge H_3 -\frac{1}{2} (B_2 - d c_1) \wedge d A_1  \right) \label{Eq:PageChargeGood2}
\end{equation}
Where $(B_2,c_1) \rightarrow (B_2 + d \Lambda_1, c_1 + \Lambda_1)$.  That this operator is gauge invariant and topological on a closed manifold can be checked by extending it to $5d$\footnote{For further arguments on its topological nature one can formulate similar considerations as the ones presented in Appendix A of \cite{GarciaEtxebarria:2022jky}.}. This new operator acts non-invertibly in sectors with either $\int H_3 = m$ or $\int dA_1 = m$ fluxes. The argument is analog to the previous case and implies the vanishing of the partition function unless
$$\alpha m \in 2 \mathbb{Z}.$$ We note that the action of this operator is hence not equivalent to \ref{Eq:TopOpIIA1} for which $\alpha$ has to satisfy a less stringent condition in terms of $m$: \ref{Eq:Condition}\footnote{For the interest of simplicity we have avoided discussing the 11d supergravity topological operator in this approach. In that case one must compare, 
\begin{equation}
    U_\alpha^{\prime}(\Sigma_7) = \int D c_2 \Big|_{\Sigma_7} \text{exp} \left( 2 \pi i \alpha \int_{\Sigma_7} \star_{11} F_4 - (A_3 - d c_2) \wedge F_4   \right) \label{Eq:PageChargeGood}
\end{equation}
and \ref{Eq:TopoDefectMtheory}. Again a factor 2 appears which implies that the two operators are not completely equivalent.}.
A further difference is that this operator admits a straightforward generalization to arbitrary Page charges. Let us write for completeness the one associated with the current in Eq. \ref{Eq:ImprovedF6}:
\begin{equation}
    \hat{U}_\alpha(\Sigma_6) = \int D c_1 D c_2 \Big|_{\Sigma_6} \text{exp} \left( 2 \pi i \alpha \int_{\Sigma_6} \Tilde{F}_6 - \frac{1}{2} (A_3 - d c_2) \wedge H_3 -\frac{1}{2} (B_2 - d c_1) \wedge d A_3  \right) \label{Eq:PageChargeGood2}
\end{equation}
Gauge invariance is a bit trickier since $dA_3$ transforms under $A_1$ gauge transformations as $dA_3 \rightarrow dA_3 - d \lambda_0 \wedge H_3$ but can be checked to hold in closed manifolds.

\section{Discussion} \label{Sec:Discussion}

In this note we have started the exploration of non-invertible symmetries in supergravity, the low energy limit of M-theory and String Theory. We have found that, in each case studied, a $U(1)$ higher form symmetry that seemed broken by Chern-Simons (or modified Bianchi Identities) can be recovered in the form of a non-invertible topological operators for each rational value of $\alpha \in [0,1)$. We have described how these operators can be constructed explicitly by higher gauging discrete subgroups of the invertible symmetries of the theory. This construction automatically implies their topological nature and allowed us to deduce their action on the different brane probes of the theory. Finally, we have constructed alternative topological operators by using Stueckelberg-like fields. This approach has the advantage of being defined for any irrational $\alpha$. Another advantage is that it admits a straightforward generalization to any Page charge, which is unclear how to do for the first method. We conclude in this section by making some comments on the applications of these symmetries, particularly in the context of the Swampland program \cite{Vafa:2005ui,Agmon:2022thq}.

An interesting application of non-invertible symmetries of this kind is that they require the existence of auxiliary symmetries that need to be gauged for the non-invertible topological operator to exist. This gives rise to a hierarchy between the scales of symmetry breaking of the different symmetries of the theory\footnote{This is clear for the operators constructed in \cref{Sec:Supergravity}, while it is less clear for the ones in \cref{Sec:DiffAproach}.}. Consider the operator in \ref{Eq:TopoDefectMtheory} for instance. The existence of the 3-form non-invertible symmetry requires the existence of an exact 6-form magnetic symmetry. This implies a hierarchy between the energy at which these symmetries are broken in a UV-completion.
\begin{equation}
        E(\Gamma_{\mathbb{Q}}^{(3)}) \leq E( U(1)_m^{(6)})
\end{equation}
In particular, if one assumes that $\Gamma_{\mathbb{Q}}^{(3)}$,  $U(1)_m^{(6)}$ are broken explicitly by including dynamical objects electrically charged under them, M2- and M5-branes, respectively, one concludes the following relation between their tension,
\begin{equation}
    T_{M2}^{1/3} \lesssim T_{M5}^{1/6}
\end{equation}
which is true up to an order 1 factor with the values $T_{M2} = (2 \pi)^{-2} l_p^{-3}, \, T_{M5} = (2 \pi)^{-5} l_p^{-6}$\footnote{Note that a similar relation is enforced by the 2-group structure.}. Similar considerations apply to the operator in \ref{Eq:TopOpIIA1}. In that case, to build $\Gamma^{(5)}_{\mathbb{Q}}$ defects we need to gauge a subgroup of $U(1)_m^{(6)} \times U(1)_m^{(7)}$ which implies,
\begin{equation}
      E(\Gamma_{\mathbb{Q}}^{(3)}) \leq \text{min} \left(E( U(1)_m^{(6)}),E( U(1)_m^{(7)}) \right)
\end{equation}
If we again assume that the symmetries are only broken by the presence of the minimal branes charged under them we have,
\begin{equation}
    T_{D4}^{1/5} \lesssim \text{min} \left(  T_{NS5}^{1/6},T_{D6}^{1/7}  \right)
\end{equation}
which again is fulfilled in Type IIA String Theory by the NS5-brane. A similar relation, which is again satisfied, applies to the topological operator of Type IIB String Theory. The assumption that the symmetries are only broken by their coupling to their minimal objects is not true in either M-theory or Type II String Theory, so these relations needed not hold. It is still amusing to see that they are indeed satisfied.

Let us now elaborate a bit on the relation between these symmetries and the completeness hypothesis. The two simplest mechanisms by which a putative UV completion breaks higher-form symmetries of the low energy theory is by the presence of Chern-Simons couplings and the introduction of dynamical objects \cite{Montero:2017yja,Heidenreich:2020pkc,Heidenreich:2021xpr}. In this work we have argued that Chern-Simons terms are generically not enough to effectively break the symmetries, making the case for the need of adding dynamical objects. This makes more clear than ever the connection between having a complete spectrum (The Completeness Hypothesis) and the absence of generalized global symmetries.

We finish with an outlook of the future work.
\begin{itemize}
    \item We have seen two different approaches to finding non-invertible symmetries for non-conservation equations. It would be very interesting to better understand the connection between the two approaches. In particular, it would help if we understood how to construct the operators in \cref{Sec:DiffAproach} explicitly, maybe from higher gauging.
    \item Relatedly, we have left for future work the understanding of whether more ``rational valued" non-invertible operators can be built in Type II supergravity by further gauging the non-invertible symmetries. 
    \item It would be very interesting to elaborate on the actions of the topological operators in the different probe branes. We expect very rich fusion rules, in a similar spirit to \cite{Choi:2022fgx}. 
    \item We expect these results to have many applications and generalizations in string compactifications. In particular, it would be very interesting to see if our methods can be generalized to compactifications with generalized $\theta$-terms, as studied in \cite{Grimm:2022xmj}.
\end{itemize}

\subsection*{Acknowledgments}

I am pleased to thank Jeremías Aguilera, Riccardo Argurio, Shani Meynet, Miguel Montero and Damian van de Heisteeg for related and insightful discussions. I would also like to thank Shani Meynet, Antoine Pasternak, Valdo Tatitscheff and specially Riccardo Argurio and Iñaki García-Etxebarria, for taking time during Christmas to read this manuscript and share important comments. Finally, I want to thank Raquel for her support. This research is supported by a Margarita Salas award CA1/RSUE/2021-00738 from the Plan de Recuperación, Transformación y Resiliencia of the Ministerio de Universidades, UAM and Harvard.

\appendix

\section{The 7d $\mathbb{Z}_N$ TQFT} \label{AppTQFT}
In this appendix we define the 7d TQFT with 3-form $\mathbb{Z}_N$ symmmetry and anomaly as needed to render the 11d supergravity defects gauge invariant. We start by briefly reviewing the 3d $A^{n,p}$ theory. Consider a smooth deformation of the would-be topological defect in \ref{Eq:BadOpeNaive} with $\alpha = p/N$. The operators fails to be topological due to the equation of motion, which gives it a phase,
\begin{equation}
    \text{exp} \left( \frac{i \pi p }{N} \int_{M_4} F_2 \wedge F_2   \right) \label{Eq:AppPhase}
\end{equation}
One looks for a TQFT that can cancel this phase. This is precisely what the $\mathcal{A}^{N,p}[B_2]$ theory does. Indeed it is defined to have 1-form symmetry $\mathbb{Z}_N^{(1)}$ and anomaly given by inflow as,
\begin{equation}
    S_{3}^{(N,p)} = - i \pi p N \int_{M_4} B_2 \wedge B_2 
\end{equation}
 Where $B_2$ is a $\mathbb{Z}^{(1)}_N$ background gauge field with holonomies in $\mathbb{Z}/N$. We may now identify the background $B_2 = F_2/N$ to cancel the phase \ref{Eq:AppPhase} so that,
 \begin{equation}
      U_{\frac{p}{N}} (\Sigma_3) \times A^{(N,p)} \left[ \frac{F_2}{N} \right] \label{Eq:GeneralOp}
 \end{equation}
is topological and gauge invariant, as reviewed in the main text. For further details on how to define the $A^{(N,p)}[B_2]$ theory the reader may check \cite{Hsin:2018vcg,Choi:2022jqy}. Consider now the case of 11d supergravity and the non-conservation equation in \ref{Eq:EOMMtheory}. The naive operator is not topological, as it picks up a phase
\begin{equation}
    \text{exp} \left( \frac{i \pi p }{N} \int_{M_8} F_4 \wedge F_4   \right) \label{Eq:AppPhaseM}
\end{equation}
In analogy with the discussion above we define a 7d TQFT $\mathcal{A}^{(N,p)}_7[B_4]$ with 3-form $\mathbb{Z}^{(3)}_N$ symmmetry and anomaly characterized by inflow as,
\begin{equation}
    S_{7}^{(N,p)} = - i \pi p N \int_{M_8} B_4 \wedge B_4 
\end{equation}
Where $B_4$ is a $\mathbb{Z}^{(3)}_N$ background gauge field with holonomies in $\mathbb{Z}/N$. 

\section{Explicit construction of the non-invertible defects by half gauging} \label{Sec:AppConstruction}
In this appendix we construct the rational valued non-invertible defects introduced in the main text by using the technique of higher gauging \cite{Roumpedakis:2022aik}. Consider first the non-invertible topological operator introduced for 11d supergravity \ref{Eq:TopoDefectMtheory}. We choose an auxiliary manifold $\Sigma_8$ such that $\Sigma_7 = \partial \Sigma_8$ and gauge a $\mathbb{Z}^{(6)}_N$ subgroup of the magnetic symmetry $U(1)^{(6)}$ with appropriate discrete torsion. This gauging is described by adding the following terms to the path integral,
\begin{equation}
    \delta S = 2 \pi i \int_{\Sigma_8} N b_4 \wedge d \hat{c}_3 + b_4 \wedge F_4 + \frac{N p^{\prime}}{2} b_4 \wedge b_4
\end{equation}
Where $p p^{\prime} = 1$ mod $N$. Let us unpack a bit the expression above. The second term describes the coupling of the $U(1)^{(6)}$ current to a a background $b_4$ in  $\Sigma_8$. The first term is a coupling to a $U(1)$ Lagrange multiplier gauge field $\hat{c}_3$ whose job is to restrict the holonomy of $b_4$ so that it is effectively a $\mathbb{Z}_N$ gauge field. The third term is a discrete torsion that one may always add. The equation of motion for $b_4$ is $N d\hat{c}_3 + F_4 + N p^{\prime} b_4 = 0$. Making $b_4$ dynamical implements the gauging. Consider a closed $\Sigma_8$. If we use the equation of motion and remove terms that are multiples of $2\pi i$, we see that our gauging precisely cancels the phase in \ref{Eq:AppPhaseM}, as needed. If we instead take $p^{\prime} = 1$ and $\Sigma_8$ manifold with boundary $\partial \Sigma_8 = \Sigma_7$ we explicitly find,
\begin{equation}
    \delta S = - i \pi N \int_{M_8} b_4 \wedge b_4  + \int_{\partial \Sigma_7} \mathcal{A}^{(N,1)}_7[B_4]
\end{equation}
We thus conclude that the gauging above precisely generates $A_7^{(N,p)}$ in $ \Sigma_7 = \partial \Sigma_8$. This explicit construction of the defect is a further check of its topological nature. An important point for this gauging to work is that the theory must be self-dual under it, as emphasized in \cite{Choi:2022fgx}. In particular, if one $p$-gauges a discrete $q$-form symmetry in a d-dimensional theory, for the symmetries to be the same before and after the gauging the following relation must hold,
\begin{equation}
    q = (d+p-2)/2
\end{equation}
In the case at hand, $q =6, \, p = 3$ and $d=11$, the relation is fulfilled.

Consider now the topological defect introduced for type IIA supergravity \ref{Eq:TopOpIIA1}. The classical phase that one whises to cancel is now,
\begin{equation}
    \text{exp} \left( \frac{2 \pi i p }{N} \int_{M_5} dA_1 \wedge H_3   \right) \label{Eq:AppPhase2IIA}
\end{equation}
We have to gauge a $\mathbb{Z}^{(6)}_N \times \mathbb{Z}^{(7)}_N$ subgroup of the magnetic symmetry $U(1)^{(6)}_m \times U(1)^{(7)}_m$ in a 5-dimensional manifold $\Sigma_5$, so we introduce two background fields $b_2$ and $b_3$, respectively. We also introduce Lagrange multipliers $\hat{c}_1, \hat{v}_2$ enforcing $\mathbb{Z}_N$ holonomies and a discrete torsion term. The mixed gauging is hence implemented by adding the following term to the path integral,
\begin{equation}
    \delta S = 2 \pi i \int_{\Sigma_5} N p^{\prime} b_3 \wedge b_2 + N b_3 \wedge d \hat{c}_1 + N b_2 \wedge d \hat{v}_2 - b_3 \wedge d A_1 - b_2 \wedge H_3
\end{equation}
Direct computation in a closed manifold using the equations of motion of $b_2, b_3$ shows that one precisely cancels the phase in \ref{Eq:AppPhase2IIA}. In a manifold with boundary and with $p=1$ one recovers the TQFT in \ref{Eq:TopOpIIA1} as expected. This gauging allows us to generalize the topological defect to arbitrary $p$. Note that in this gauging the quantum symmetries are interchanged, in the sense that the quantum symmetry from gauging $\mathbb{Z}^{(6)}_N$ becomes part of $U(1)^{(7)}_m$ after gauging and viceversa. This ensures that the symmetry before and after the gauging is the same. A similar construction applies with minor modifications to the Type IIB defect in \ref{Eq:TopOpIIB1}.

\bibliographystyle{JHEP}
\bibliography{Bib}

\providecommand{\href}[2]{#2}\begingroup\raggedright\begin{thebibliography}{10}

\bibitem{Gaiotto:2014kfa}
D.~Gaiotto, A.~Kapustin, N.~Seiberg, and B.~Willett, {\it {Generalized Global
  Symmetries}},  {\em JHEP} {\bf 02} (2015) 172,
  [\href{http://arxiv.org/abs/1412.5148}{{\tt arXiv:1412.5148}}].

\bibitem{Rudelius:2020orz}
T.~Rudelius and S.-H. Shao, {\it {Topological Operators and Completeness of
  Spectrum in Discrete Gauge Theories}},  {\em JHEP} {\bf 12} (2020) 172,
  [\href{http://arxiv.org/abs/2006.10052}{{\tt arXiv:2006.10052}}].

\bibitem{Heidenreich:2021xpr}
B.~Heidenreich, J.~McNamara, M.~Montero, M.~Reece, T.~Rudelius, and
  I.~Valenzuela, {\it {Non-invertible global symmetries and completeness of the
  spectrum}},  {\em JHEP} {\bf 09} (2021) 203,
  [\href{http://arxiv.org/abs/2104.07036}{{\tt arXiv:2104.07036}}].

\bibitem{Nguyen:2021yld}
M.~Nguyen, Y.~Tanizaki, and M.~\"Unsal, {\it {Semi-Abelian gauge theories,
  non-invertible symmetries, and string tensions beyond $N$-ality}},  {\em
  JHEP} {\bf 03} (2021) 238, [\href{http://arxiv.org/abs/2101.02227}{{\tt
  arXiv:2101.02227}}].

\bibitem{Koide:2021zxj}
M.~Koide, Y.~Nagoya, and S.~Yamaguchi, {\it {Non-invertible topological defects
  in 4-dimensional $\mathbb {Z}_2$ pure lattice gauge theory}},  {\em PTEP}
  {\bf 2022} (2022), no.~1 013B03, [\href{http://arxiv.org/abs/2109.05992}{{\tt
  arXiv:2109.05992}}].

\bibitem{Choi:2021kmx}
Y.~Choi, C.~Cordova, P.-S. Hsin, H.~T. Lam, and S.-H. Shao, {\it
  {Non-Invertible Duality Defects in 3+1 Dimensions}},
  \href{http://arxiv.org/abs/2111.01139}{{\tt arXiv:2111.01139}}.

\bibitem{Kaidi:2021xfk}
J.~Kaidi, K.~Ohmori, and Y.~Zheng, {\it {Kramers-Wannier-like Duality Defects
  in (3+1)D Gauge Theories}},  {\em Phys. Rev. Lett.} {\bf 128} (2022), no.~11
  111601, [\href{http://arxiv.org/abs/2111.01141}{{\tt arXiv:2111.01141}}].

\bibitem{Cordova:2022rer}
C.~Cordova, K.~Ohmori, and T.~Rudelius, {\it {Generalized Symmetry Breaking
  Scales and Weak Gravity Conjectures}},
  \href{http://arxiv.org/abs/2202.05866}{{\tt arXiv:2202.05866}}.

\bibitem{Benini:2022hzx}
F.~Benini, C.~Copetti, and L.~Di~Pietro, {\it {Factorization and global
  symmetries in holography}},  \href{http://arxiv.org/abs/2203.09537}{{\tt
  arXiv:2203.09537}}.

\bibitem{Roumpedakis:2022aik}
K.~Roumpedakis, S.~Seifnashri, and S.-H. Shao, {\it {Higher Gauging and
  Non-invertible Condensation Defects}},
  \href{http://arxiv.org/abs/2204.02407}{{\tt arXiv:2204.02407}}.

\bibitem{Bhardwaj:2022yxj}
L.~Bhardwaj, L.~Bottini, S.~Schafer-Nameki, and A.~Tiwari, {\it {Non-Invertible
  Higher-Categorical Symmetries}},  \href{http://arxiv.org/abs/2204.06564}{{\tt
  arXiv:2204.06564}}.

\bibitem{Arias-Tamargo:2022nlf}
G.~Arias-Tamargo and D.~Rodriguez-Gomez, {\it {Non-Invertible Symmetries from
  Discrete Gauging and Completeness of the Spectrum}},
  \href{http://arxiv.org/abs/2204.07523}{{\tt arXiv:2204.07523}}.

\bibitem{Hayashi:2022fkw}
Y.~Hayashi and Y.~Tanizaki, {\it {Non-invertible self-duality defects of
  Cardy-Rabinovici model and mixed gravitational anomaly}},
  \href{http://arxiv.org/abs/2204.07440}{{\tt arXiv:2204.07440}}.

\bibitem{Choi:2022zal}
Y.~Choi, C.~Cordova, P.-S. Hsin, H.~T. Lam, and S.-H. Shao, {\it
  {Non-invertible Condensation, Duality, and Triality Defects in 3+1
  Dimensions}},  \href{http://arxiv.org/abs/2204.09025}{{\tt
  arXiv:2204.09025}}.

\bibitem{Kaidi:2022uux}
J.~Kaidi, G.~Zafrir, and Y.~Zheng, {\it {Non-Invertible Symmetries of
  $\mathcal{N}=4$ SYM and Twisted Compactification}},
  \href{http://arxiv.org/abs/2205.01104}{{\tt arXiv:2205.01104}}.

\bibitem{Choi:2022jqy}
Y.~Choi, H.~T. Lam, and S.-H. Shao, {\it {Non-invertible Global Symmetries in
  the Standard Model}},  \href{http://arxiv.org/abs/2205.05086}{{\tt
  arXiv:2205.05086}}.

\bibitem{Cordova:2022ieu}
C.~Cordova and K.~Ohmori, {\it {Non-Invertible Chiral Symmetry and Exponential
  Hierarchies}},  \href{http://arxiv.org/abs/2205.06243}{{\tt
  arXiv:2205.06243}}.

\bibitem{Antinucci:2022eat}
A.~Antinucci, G.~Galati, and G.~Rizi, {\it {On Continuous 2-Category Symmetries
  and Yang-Mills Theory}},  \href{http://arxiv.org/abs/2206.05646}{{\tt
  arXiv:2206.05646}}.

\bibitem{Bashmakov:2022jtl}
V.~Bashmakov, M.~Del~Zotto, and A.~Hasan, {\it {On the 6d Origin of
  Non-invertible Symmetries in 4d}},
  \href{http://arxiv.org/abs/2206.07073}{{\tt arXiv:2206.07073}}.

\bibitem{Damia:2022rxw}
J.~Aguilera~Damia, R.~Argurio, and L.~Tizzano, {\it {Continuous Generalized
  Symmetries in Three Dimensions}},
  \href{http://arxiv.org/abs/2206.14093}{{\tt arXiv:2206.14093}}.

\bibitem{Damia:2022bcd}
J.~A. Damia, R.~Argurio, and E.~Garcia-Valdecasas, {\it {Non-Invertible Defects
  in 5d, Boundaries and Holography}},
  \href{http://arxiv.org/abs/2207.02831}{{\tt arXiv:2207.02831}}.

\bibitem{Choi:2022rfe}
Y.~Choi, H.~T. Lam, and S.-H. Shao, {\it {Non-invertible Time-reversal
  Symmetry}},  \href{http://arxiv.org/abs/2208.04331}{{\tt arXiv:2208.04331}}.

\bibitem{Bhardwaj:2022lsg}
L.~Bhardwaj, S.~Schafer-Nameki, and J.~Wu, {\it {Universal Non-Invertible
  Symmetries}},  {\em Fortsch. Phys.} {\bf 70} (2022), no.~11 2200143,
  [\href{http://arxiv.org/abs/2208.05973}{{\tt arXiv:2208.05973}}].

\bibitem{Bartsch:2022mpm}
T.~Bartsch, M.~Bullimore, A.~E.~V. Ferrari, and J.~Pearson, {\it
  {Non-invertible Symmetries and Higher Representation Theory I}},
  \href{http://arxiv.org/abs/2208.05993}{{\tt arXiv:2208.05993}}.

\bibitem{Lin:2022xod}
L.~Lin, D.~G. Robbins, and E.~Sharpe, {\it {Decomposition, Condensation
  Defects, and Fusion}},  {\em Fortsch. Phys.} {\bf 70} (2022), no.~11 2200130,
  [\href{http://arxiv.org/abs/2208.05982}{{\tt arXiv:2208.05982}}].

\bibitem{GarciaEtxebarria:2022vzq}
I.~n. Garc\'\i{}a~Etxebarria, {\it {Branes and Non-Invertible Symmetries}},
  {\em Fortsch. Phys.} {\bf 70} (2022), no.~11 2200154,
  [\href{http://arxiv.org/abs/2208.07508}{{\tt arXiv:2208.07508}}].

\bibitem{Apruzzi:2022rei}
F.~Apruzzi, I.~Bah, F.~Bonetti, and S.~Schafer-Nameki, {\it {Non-Invertible
  Symmetries from Holography and Branes}},
  \href{http://arxiv.org/abs/2208.07373}{{\tt arXiv:2208.07373}}.

\bibitem{Heckman:2022muc}
J.~J. Heckman, M.~H\"ubner, E.~Torres, and H.~Y. Zhang, {\it {The Branes Behind
  Generalized Symmetry Operators}},
  \href{http://arxiv.org/abs/2209.03343}{{\tt arXiv:2209.03343}}.

\bibitem{Freed:2022qnc}
D.~S. Freed, G.~W. Moore, and C.~Teleman, {\it {Topological symmetry in quantum
  field theory}},  \href{http://arxiv.org/abs/2209.07471}{{\tt
  arXiv:2209.07471}}.

\bibitem{Niro:2022ctq}
P.~Niro, K.~Roumpedakis, and O.~Sela, {\it {Exploring Non-Invertible Symmetries
  in Free Theories}},  \href{http://arxiv.org/abs/2209.11166}{{\tt
  arXiv:2209.11166}}.

\bibitem{Kaidi:2022cpf}
J.~Kaidi, K.~Ohmori, and Y.~Zheng, {\it {Symmetry TFTs for Non-Invertible
  Defects}},  \href{http://arxiv.org/abs/2209.11062}{{\tt arXiv:2209.11062}}.

\bibitem{Mekareeya:2022spm}
N.~Mekareeya and M.~Sacchi, {\it {Mixed Anomalies, Two-groups, Non-Invertible
  Symmetries, and 3d Superconformal Indices}},
  \href{http://arxiv.org/abs/2210.02466}{{\tt arXiv:2210.02466}}.

\bibitem{Antinucci:2022vyk}
A.~Antinucci, F.~Benini, C.~Copetti, G.~Galati, and G.~Rizi, {\it {The
  holography of non-invertible self-duality symmetries}},
  \href{http://arxiv.org/abs/2210.09146}{{\tt arXiv:2210.09146}}.

\bibitem{Chen:2022cyw}
S.~Chen and Y.~Tanizaki, {\it {Solitonic symmetry beyond homotopy:
  invertibility from bordism and non-invertibility from TQFT}},
  \href{http://arxiv.org/abs/2210.13780}{{\tt arXiv:2210.13780}}.

\bibitem{Bashmakov:2022uek}
V.~Bashmakov, M.~Del~Zotto, A.~Hasan, and J.~Kaidi, {\it {Non-invertible
  Symmetries of Class $\mathcal{S}$ Theories}},
  \href{http://arxiv.org/abs/2211.05138}{{\tt arXiv:2211.05138}}.

\bibitem{Karasik:2022kkq}
A.~Karasik, {\it {On anomalies and gauging of U(1) non-invertible symmetries in
  4d QED}},  \href{http://arxiv.org/abs/2211.05802}{{\tt arXiv:2211.05802}}.

\bibitem{Cordova:2022fhg}
C.~Cordova, S.~Hong, S.~Koren, and K.~Ohmori, {\it {Neutrino Masses from
  Generalized Symmetry Breaking}},  \href{http://arxiv.org/abs/2211.07639}{{\tt
  arXiv:2211.07639}}.

\bibitem{GarciaEtxebarria:2022jky}
I.~n. Garc\'\i{}a~Etxebarria and N.~Iqbal, {\it {A Goldstone theorem for
  continuous non-invertible symmetries}},
  \href{http://arxiv.org/abs/2211.09570}{{\tt arXiv:2211.09570}}.

\bibitem{Verlinde:1988sn}
E.~P. Verlinde, {\it {Fusion Rules and Modular Transformations in 2D Conformal
  Field Theory}},  {\em Nucl. Phys. B} {\bf 300} (1988) 360--376.

\bibitem{Petkova:2000ip}
V.~B. Petkova and J.~B. Zuber, {\it {Generalized twisted partition functions}},
   {\em Phys. Lett. B} {\bf 504} (2001) 157--164,
  [\href{http://arxiv.org/abs/hep-th/0011021}{{\tt hep-th/0011021}}].

\bibitem{Fuchs:2002cm}
J.~Fuchs, I.~Runkel, and C.~Schweigert, {\it {TFT construction of RCFT
  correlators 1. Partition functions}},  {\em Nucl. Phys. B} {\bf 646} (2002)
  353--497, [\href{http://arxiv.org/abs/hep-th/0204148}{{\tt hep-th/0204148}}].

\bibitem{Frohlich:2004ef}
J.~Frohlich, J.~Fuchs, I.~Runkel, and C.~Schweigert, {\it {Kramers-Wannier
  duality from conformal defects}},  {\em Phys. Rev. Lett.} {\bf 93} (2004)
  070601, [\href{http://arxiv.org/abs/cond-mat/0404051}{{\tt
  cond-mat/0404051}}].

\bibitem{Bhardwaj:2017xup}
L.~Bhardwaj and Y.~Tachikawa, {\it {On finite symmetries and their gauging in
  two dimensions}},  {\em JHEP} {\bf 03} (2018) 189,
  [\href{http://arxiv.org/abs/1704.02330}{{\tt arXiv:1704.02330}}].

\bibitem{Tachikawa:2017gyf}
Y.~Tachikawa, {\it {On gauging finite subgroups}},  {\em SciPost Phys.} {\bf 8}
  (2020), no.~1 015, [\href{http://arxiv.org/abs/1712.09542}{{\tt
  arXiv:1712.09542}}].

\bibitem{Chang:2018iay}
C.-M. Chang, Y.-H. Lin, S.-H. Shao, Y.~Wang, and X.~Yin, {\it {Topological
  Defect Lines and Renormalization Group Flows in Two Dimensions}},  {\em JHEP}
  {\bf 01} (2019) 026, [\href{http://arxiv.org/abs/1802.04445}{{\tt
  arXiv:1802.04445}}].

\bibitem{Thorngren:2019iar}
R.~Thorngren and Y.~Wang, {\it {Fusion Category Symmetry I: Anomaly In-Flow and
  Gapped Phases}},  \href{http://arxiv.org/abs/1912.02817}{{\tt
  arXiv:1912.02817}}.

\bibitem{Gaiotto:2020iye}
D.~Gaiotto and J.~Kulp, {\it {Orbifold groupoids}},  {\em JHEP} {\bf 02} (2021)
  132, [\href{http://arxiv.org/abs/2008.05960}{{\tt arXiv:2008.05960}}].

\bibitem{Komargodski:2020mxz}
Z.~Komargodski, K.~Ohmori, K.~Roumpedakis, and S.~Seifnashri, {\it {Symmetries
  and strings of adjoint QCD$_{2}$}},  {\em JHEP} {\bf 03} (2021) 103,
  [\href{http://arxiv.org/abs/2008.07567}{{\tt arXiv:2008.07567}}].

\bibitem{Nguyen:2021naa}
M.~Nguyen, Y.~Tanizaki, and M.~\"Unsal, {\it {Noninvertible 1-form symmetry and
  Casimir scaling in 2D Yang-Mills theory}},  {\em Phys. Rev. D} {\bf 104}
  (2021), no.~6 065003, [\href{http://arxiv.org/abs/2104.01824}{{\tt
  arXiv:2104.01824}}].

\bibitem{Thorngren:2021yso}
R.~Thorngren and Y.~Wang, {\it {Fusion Category Symmetry II: Categoriosities at
  $c$ = 1 and Beyond}},  \href{http://arxiv.org/abs/2106.12577}{{\tt
  arXiv:2106.12577}}.

\bibitem{Montero:2017yja}
M.~Montero, A.~M. Uranga, and I.~Valenzuela, {\it {A Chern-Simons Pandemic}},
  {\em JHEP} {\bf 07} (2017) 123, [\href{http://arxiv.org/abs/1702.06147}{{\tt
  arXiv:1702.06147}}].

\bibitem{Banks:1988yz}
T.~Banks and L.~J. Dixon, {\it {Constraints on String Vacua with Space-Time
  Supersymmetry}},  {\em Nucl. Phys. B} {\bf 307} (1988) 93--108.

\bibitem{Banks:2010zn}
T.~Banks and N.~Seiberg, {\it {Symmetries and Strings in Field Theory and
  Gravity}},  {\em Phys. Rev. D} {\bf 83} (2011) 084019,
  [\href{http://arxiv.org/abs/1011.5120}{{\tt arXiv:1011.5120}}].

\bibitem{Harlow:2018tng}
D.~Harlow and H.~Ooguri, {\it {Symmetries in quantum field theory and quantum
  gravity}},  {\em Commun. Math. Phys.} {\bf 383} (2021), no.~3 1669--1804,
  [\href{http://arxiv.org/abs/1810.05338}{{\tt arXiv:1810.05338}}].

\bibitem{Harlow:2020bee}
D.~Harlow and E.~Shaghoulian, {\it {Global symmetry, Euclidean gravity, and the
  black hole information problem}},  {\em JHEP} {\bf 04} (2021) 175,
  [\href{http://arxiv.org/abs/2010.10539}{{\tt arXiv:2010.10539}}].

\bibitem{Chen:2020ojn}
Y.~Chen and H.~W. Lin, {\it {Signatures of global symmetry violation in
  relative entropies and replica wormholes}},  {\em JHEP} {\bf 03} (2021) 040,
  [\href{http://arxiv.org/abs/2011.06005}{{\tt arXiv:2011.06005}}].

\bibitem{Hsin:2020mfa}
P.-S. Hsin, L.~V. Iliesiu, and Z.~Yang, {\it {A violation of global symmetries
  from replica wormholes and the fate of black hole remnants}},  {\em Class.
  Quant. Grav.} {\bf 38} (2021), no.~19 194004,
  [\href{http://arxiv.org/abs/2011.09444}{{\tt arXiv:2011.09444}}].

\bibitem{Sasieta:2022ksu}
M.~Sasieta, {\it {Wormholes from heavy operator statistics in AdS/CFT}},
  \href{http://arxiv.org/abs/2211.11794}{{\tt arXiv:2211.11794}}.

\bibitem{Bah:2022uyz}
I.~Bah, Y.~Chen, and J.~Maldacena, {\it {Estimating global charge violating
  amplitudes from wormholes}},  \href{http://arxiv.org/abs/2212.08668}{{\tt
  arXiv:2212.08668}}.

\bibitem{Heidenreich:2020pkc}
B.~Heidenreich, J.~McNamara, M.~Montero, M.~Reece, T.~Rudelius, and
  I.~Valenzuela, {\it {Chern-Weil global symmetries and how quantum gravity
  avoids them}},  {\em JHEP} {\bf 11} (2021) 053,
  [\href{http://arxiv.org/abs/2012.00009}{{\tt arXiv:2012.00009}}].

\bibitem{Marolf:2000cb}
D.~Marolf, {\it {Chern-Simons terms and the three notions of charge}},  in {\em
  {International Conference on Quantization, Gauge Theory, and Strings:
  Conference Dedicated to the Memory of Professor Efim Fradkin}}, pp.~312--320,
  6, 2000.
\newblock \href{http://arxiv.org/abs/hep-th/0006117}{{\tt hep-th/0006117}}.

\bibitem{Petersson:2007sc}
C.~Petersson, {\it {Superpotentials From Stringy Instantons Without
  Orientifolds}},  {\em JHEP} {\bf 05} (2008) 078,
  [\href{http://arxiv.org/abs/0711.1837}{{\tt arXiv:0711.1837}}].

\bibitem{Seiberg:1999vs}
N.~Seiberg and E.~Witten, {\it {String theory and noncommutative geometry}},
  {\em JHEP} {\bf 09} (1999) 032,
  [\href{http://arxiv.org/abs/hep-th/9908142}{{\tt hep-th/9908142}}].

\bibitem{Hsin:2018vcg}
P.-S. Hsin, H.~T. Lam, and N.~Seiberg, {\it {Comments on One-Form Global
  Symmetries and Their Gauging in 3d and 4d}},  {\em SciPost Phys.} {\bf 6}
  (2019), no.~3 039, [\href{http://arxiv.org/abs/1812.04716}{{\tt
  arXiv:1812.04716}}].

\bibitem{Heckman:2017uxe}
J.~J. Heckman and L.~Tizzano, {\it {6D Fractional Quantum Hall Effect}},  {\em
  JHEP} {\bf 05} (2018) 120, [\href{http://arxiv.org/abs/1708.02250}{{\tt
  arXiv:1708.02250}}].

\bibitem{Choi:2022fgx}
Y.~Choi, H.~T. Lam, and S.-H. Shao, {\it {Non-invertible Gauss Law and
  Axions}},  \href{http://arxiv.org/abs/2212.04499}{{\tt arXiv:2212.04499}}.

\bibitem{Becker:2006dvp}
K.~Becker, M.~Becker, and J.~H. Schwarz, {\em {String theory and M-theory: A
  modern introduction}}.
\newblock Cambridge University Press, 12, 2006.

\bibitem{Yokokura:2022alv}
R.~Yokokura, {\it {Non-invertible symmetries in axion electrodynamics}},
  \href{http://arxiv.org/abs/2212.05001}{{\tt arXiv:2212.05001}}.

\bibitem{Vafa:2005ui}
C.~Vafa, {\it {The String landscape and the swampland}},
  \href{http://arxiv.org/abs/hep-th/0509212}{{\tt hep-th/0509212}}.

\bibitem{Agmon:2022thq}
N.~B. Agmon, A.~Bedroya, M.~J. Kang, and C.~Vafa, {\it {Lectures on the string
  landscape and the Swampland}},  \href{http://arxiv.org/abs/2212.06187}{{\tt
  arXiv:2212.06187}}.

\bibitem{Grimm:2022xmj}
T.~W. Grimm, S.~Lanza, and T.~van Vuren, {\it {Global symmetry-breaking and
  generalized theta-terms in Type IIB EFTs}},
  \href{http://arxiv.org/abs/2211.11769}{{\tt arXiv:2211.11769}}.

\end{thebibliography}\endgroup


\end{document}